\documentclass[traditabstract]{aa}

\usepackage{txfonts}
\usepackage{times}
\usepackage{epsf}
\usepackage[dvips]{epsfig}
\usepackage{natbib}

\begin{document}

\title{Leo IV \& V -- A possible dwarf galaxy pair ?}
\titlerunning{Leo IV \& V} 

\author{M. Bla\~{n}a \and M. Fellhauer \and R. Smith}
\institute{Departamento de Astronom\'{i}a, Universidad de
  Concepci\'{o}n, Casilla 160-C, Concepci\'{o}n, 
  Chile, \email{mblana@udec.cl}}
\authorrunning{M. Bla\~{n}a et al.}

\date{Received XXX / Accepted XXX}

\abstract{
  The last few years have seen the discovery of many faint and
  ultra-faint dwarf spheroidal galaxies around the Milky Way.  Among
  these is a pair of satellites called Leo~IV and Leo~V.  This pair is
  found at large distances from the Milky Way ($154$ and $175$~kpc
  respectively).  The rather small difference in radial distance, and
  the fact that they also show a close projected distance on the sky,
  has led to the idea that we might be seeing a new pair of bound
  galaxies - like the Magellanic Clouds.  In this paper we investigate
  this speculation by means of a simple integration code (confirming
  the results with full N-body simulations).  As the luminous mass of
  both faint dwarfs is far too low to allow them to be bound, we
  simulate the pair assuming extended dark matter haloes.  Our results
  show that the minimum dark matter mass required for the pair to be
  bound is rather high - ranging from $1.6 \times 10^{10}$~M$_{\odot}$
  to $5.4 \times 10^{10}$~M$_{\odot}$ (within the virial radii).
  Computing the mass of dark matter within a commonly adopted radius of
  $300$~pc shows that our models are well within the predicted range
  of dark matter content for satellites so faint.  We therefore
  conclude that it could be possible that the two galaxies constitute
  a bound pair.}  


\maketitle

\section{Introduction}
\label{sec:intro}

The last decade has seen the discovery of many new faint dwarf
spheroidal (dSph) galaxies of the Milky Way (MW) \citep[e.g.][and 
many more]{Wil05,Bel06,Zuc06a,Bel07,Wal07}.  Many of these dwarfs are
less luminous than a globular cluster (or even an open cluster) and
exhibit high velocity dispersions (given their luminous mass) 
\citep[e.g.][]{Sim07,Koc09,Geh09}.  Should these objects be in virial
equilibrium, they are the most dark matter (DM) dominated objects
known in the universe.  They would exhibit mass-to-light (M/L) ratios
of more than a thousand \citep[e.g.][]{Sim07,Fel08,Geh09}.

$\Lambda$CDM simulations \citep[e.g. Millennium II simulation
of][]{Boy09}, \citep[Via Lactea INCITE simulation of][]{Kuh08} predict
that a galaxy like our MW should be surrounded by hundreds, if not
thousands of small DM haloes which could host a dwarf galaxy.  The
discrepancy between the known number of MW satellites and these
predicted values is known as the `missing satellite problem'
\citep[e.g.][]{Kly99,Moo99}.  The discovery of new faint dwarfs in the
SDSS catalogue doubled the number of known satellites.  Extrapolating
to areas of the sky and distances not covered by the survey
\citep[e.g.][]{Kop08,Mac09, Mac10} may suggest that the missing
satellite problem is now solved.  

Amongst these ultra-faint galaxies are dwarfs which are extremely
metal-poor \citep[][and references therein]{Kir08}, some show
complex star formation histories \citep[Leo T][]{Jon08} and many of
them show unusual morphologies \citep{Col07,San09}.  Others seem to
show signs of tidal disruption \citep{Zuc06b,Fel07,Mun08}.

Signs of tidal disruption and the fact that all known dwarfs seem to
be aligned in a disc-like structure around the MW
\citep[e.g.][]{Met08,Met09} has given rise to alternative explanations 
for the existence of those galaxies.  These theories imply that most,
if not all, of the MW dwarfs are in fact no more than disrupting
star clusters or tidal dwarf galaxies hosting no DM at all
\citep{Paw11}. 

\begin{table}
  \centering
  \caption{Observational properties of Leo~IV and Leo~V.  The data is
    mainly taken from \citet{Jon10}.}
  \label{tab:obs}
  \begin{tabular}{lrr}
    Galaxy & Leo IV & Leo V  \\ 
    \hline 
    RA (J2000) & 11$^{\rm h}$ 32$^{\rm m}$ 58$^{\rm s}\!\!.6$ $\pm$
    1$^{\rm s}\!\!.6$ & 11$^{\rm h}$ 31$^{\rm m}$ 8$^{\rm s}\!\!.4$
    $\pm$ 1$^{\rm s}\!\!.6$ \\ 
    Dec (J2000) & 00$^{\rm o}$ 33' 6'' $\pm$ 54'' & 02$^{\rm o}$ 12'
    57'' $\pm$ 12'' \\  
    Dist [kpc] & 154 $\pm$ 5 & 175 $\pm$ 9 \\
    $v_{\rm GSR}$ [km\,s$^{-1}$] & 10.1 & 60.8 \\
    $L_{\rm V}$ [L$_{\odot}$] & $1.8 \pm 0.8 \times 10^{4}$ & $1.0 \pm
    0.8 \times 10^{4}$ \\
    $M_{\rm V}$ & $-5.8 \pm 0.4$ & $-5.2 \pm 0.4$ \\ 
    $r_{\rm h}$ [pc] & $128$ & $133$ \\
    $\sigma_{\rm los}$ [km\,s$^{-1}$] & $3.3 \pm 1.7$ & $2.4$--$3.7$ \\
    \hline
     \end{tabular}
\end{table}

In this paper we focus on two of these new ultra-faint dwarfs, namely
Leo~IV \citep{Bel07} and Leo~V \citep{Bel08}.  Their properties have
been studied by many authors: for Leo~IV:
\citep[][]{Mor09,Sim10,San10} and for Leo~V:
\citep[][]{Wal09a,Wal09b}.  We summarize a selection of the
observational data in Tab.~\ref{tab:obs}.  The pair of galaxies is
found at a rather large distance from the MW ($154$ and $175$~kpc
respectively).  The two galaxies are not only very close to each other
in radial distance ($22$~kpc) \citep{Mor09,Jon10} but also in
projected distance on the sky.  Their radial velocity differs only by
about $50$~km\,s$^{-1}$.  In the discovery paper of Leo~V the authors
speculate that the two dwarfs could be a bound pair similar to the
Magellanic Clouds.  In particular, the smaller dwarf, Leo~V, exhibits
deformed elongated contours \citep{Wal09a}, which could be signs of
tidal interaction.  Furthermore, there is a tentative stellar bridge
between the two satellites \citep{Jon10}, {\bf which now is more
  likely a foreground stream of the Virgo overdensity \citep{Jin12}.} 
\citet{Jon10} argued that,  
to form a bound object, the twin system would need a lot of DM - much 
more than is seen in similar faint satellites.  Nevertheless, the
authors claimed that it is highly unlikely that the two satellites are
a simple by-chance alignment.  They also rule out the possibility that
the two faint dwarfs are not galaxies at all but simple density
enhancements of a stellar stream by orbital arguments.  They conclude
that they might be a `tumbling pair' of galaxies that have fallen into
the MW together. 

With our paper we want to investigate the hypothesis of a bound pair
further.  Using a simple two-body integration method we investigate a
large part of the possible parameter space, searching for the minimal
total mass the system needs, in order to form a tightly bound pair.
We assume that both galaxies have their own DM halo, orbiting each
other.  We describe the setup of our simulations in the next section.
We then report and verify (using full N-body simulations) our results
in Sect.~\ref{sec:res} and finally discuss our findings in
Sect.~\ref{sec:conc}.  

\section{Setup}
\label{sec:setup}

\subsection{Parameter Space}
\label{sec:obscon}

\begin{table}
  \centering
  \caption{Initial positions and velocities in the adopted Cartesian
    coordinate system.  The final set of velocities assumes the same
    amount of relative tangential velocity as given by the relative
    radial velocity.} 
  \label{tab:vel}
  \begin{tabular}{crrr}
    & Leo~IV & Leo~V & $\Delta$ \\ 
    \hline
    $X$ [kpc] &  14.713 &  20.937 &  6.224 \\
    $Y$ [kpc] &  84.728 &  90.435 &  5.707 \\
    $Z$ [kpc] & 128.442 & 149.262 & 20.840 \\ 
    \hline
    $V^{r}_{X}$ [km\,s$^{-1}$] & 0.961 &  7.242 &  6.281 \\
    $V^{r}_{Y}$ [km\,s$^{-1}$] & 5.537 & 31.282 & 25.745 \\
    $V^{r}_{Z}$ [km\,s$^{-1}$] & 8.392 & 51.630 & 43.238 \\ 
    \hline
    $V^{r+t}_{X}$ [km\,s$^{-1}$] & -4.409 &  39.276 &  43.686 \\
    $V^{r+t}_{Y}$ [km\,s$^{-1}$] & 12.945 & -14.784 & -27.731 \\
    $V^{r+t}_{Z}$ [km\,s$^{-1}$] &  4.118 &  75.047 &  70.928 \\
    \hline
  \end{tabular}
\end{table}

We use the positions and distances, as reported in the discovery
papers of \citet{Bel07,Bel08}.  This enables us to transform the
positions into a Cartesian coordinate system.  The coordinates for
Leo~IV and Leo~V are shown in the first three lines in
Tab.~\ref{tab:vel}.   

Current observations only provide radial velocities for the pair.
Nothing is known so far about their proper motions.  We therefore
adopt two velocity cases in our simulations.  First, we investigate
the case that the radial velocity is the only velocity component the
galaxies have (line 4-6 in Tab.~\ref{tab:vel}) and second, we adopt an 
outward tangential velocity of the same magnitude as the radial
velocity (line 7-9 in Tab.~\ref{tab:vel}).  Here `outward
tangential' means that the velocity vector is perpendicular to the
radial velocity and points away from the other dwarf.  The latter
choice could be regarded as the worst case scenario.  With these 
assumptions we restrict the possible parameter space of initial
conditions significantly, but we are able to deduce how strong the
influence of a tentative tangential velocity is on our results.

To restrict our parameter space further we adopt just two cases for
the mass-ratio between the two satellites.  The observed absolute
magnitudes give a luminosity ratio of $1.8$.  Adopting a
mass-follows-light scenario for one of our cases, we use the same
ratio for the two DM haloes: 
\begin{equation}
  \label{eq:ratio}
  \frac{L_{\rm LeoIV}} {L_{\rm LeoV}} = \frac{1.8 \cdot 10^4} {1 \cdot
    10^4} = 1.8 = \frac{M^*_{\rm LeoIV}} {M^*_{\rm LeoV}} = \frac{M^{\rm
      DM}_{\rm LeoIV}} {M^{\rm DM}_{\rm LeoV}}.
\end{equation}
\citet{Wal09b} suggest that the faint and ultra-faint dwarfs reside
in similar DM haloes of a certain minimum mass.  Therefore, we
investigate equal mass DM haloes as the other case.

For the four cases described above we search for solutions
adopting halo concentrations of $c = 5$, $10$ and $20$, as those
are values typically adopted for dwarf galaxies \citep{Lok01}.  This
gives a total of 12 different solutions to the problem. 

The haloes are described as NFW-profiles with $c = r_{\rm vir} /
r_{\rm scale}$, $r_{\rm vir} = r_{200}$ being the virial radius in
which the density is $200$ times the critical density of the universe,
using a standard value of the Hubble constant of $H_{0} =
70$~km\,s$^{-1}$\,Mpc$^{-1}$. 

\begin{table*}
  \centering
  \caption{This table shows the minimum bound mass for each of our
    cases.  The first column gives the number of the case, the second
    is the adopted concentration of the haloes.  Then we give the mass
    of the DM halo and its virial radius for Leo~IV and Leo~V.  The
    next column gives the total mass in DM of the whole system, the
    next column shows the `ratio' by which the maximum distance
    differs between the full N-body simulation and the two-body code.
    The last column is a short explanation for the cases: rad. vel. =
    only radial velocity, rad. \& tan. vel. = radial and tangential
    velocity adopted; mass ratio 1.8 = the two haloes have a fixed
    mass ratio of 1.8; equal mass = the two haloes have the same mass}
  \label{tab:res}
  \begin{tabular}{ccccccccl}
    Case  & c & $M_{\rm DM,LeoIV}$ & $r_{\rm vir,LeoIV}$ & $M_{\rm
      DM,LeoV}$ & $r_{\rm vir,LeoV}$ & $M_{\rm tot}$ & ratio &
    Scenario \\  
    &   & [M$_{\odot}$] & [kpc] & [M$_{\odot}$] & [kpc] &
    [M$_{\odot}$] & & \\
    \hline
    0a & $\infty$ & -- & -- & -- & -- & $4.18\times 10^{9}$ & ---
    & Point masses (a) \\ 
    0b & $\infty$ & -- & -- & -- & -- & $1.47\times 10^{10}$ &
    --- & Point masses (b) \\  
    \hline
    1  &  5 & $1.34 \times 10^{10}$ & 49.00 & $7.45 \times 10^{9}$ &
    40.29 & $2.09 \times 10^{10}$ & 0.965 & rad. vel. mass
    ratio 1.8 \\  
    2  & 10 & $1.27 \times 10^{10}$ & 48.11 & $7.05 \times 10^{9}$ &
    39.55 & $1.98 \times 10^{10}$ & 0.953 & rad. vel. mass
    ratio 1.8 \\  
    3  & 20 & $1.22 \times 10^{10}$ & 47.42 & $6.75 \times 10^{9}$ &
    39.98 & $1.89 \times 10^{10}$ & 0.935 & rad. vel. mass
    ratio 1.8 \\  
    \hline
    1a &  5 & $9.05 \times 10^{9}$ & 42.99 & $9.05 \times 10^{9}$ &
    42.99 & $1.81 \times 10^{10}$ & 1.179 & rad. vel. equal
    mass \\ 
    2a & 10 & $8.55 \times 10^{9}$ & 42.18 & $8.55 \times 10^{9}$ &
    42.18 & $1.71 \times 10^{10}$ & 1.194 & rad. vel. equal
    mass \\ 
    3a & 20 & $8.30 \times 10^{9}$ & 41.76 & $8.30 \times 10^{9}$ &
    41.76 & $1.66 \times 10^{10}$ & 1.204 & rad. vel. equal
    mass \\ 
    \hline
    4  &  5 & $3.47 \times 10^{10}$ &   67.25 & $1.93 \times 10^{10}$
    & 55.28 & $5.39 \times 10^{10}$ & 0.993 & rad.\&
    tang. vel. mass ratio 1.8 \\ 
    5  & 10 & $3.11 \times 10^{10}$ & 64.83 & $1.73 \times 10^{10}$ &
    53.30 & $4.83 \times 10^{10}$ & 0.956 & rad.\&
    tang. vel. mass ratio 1.8 \\  
    6  & 20 & $2.84 \times 10^{10}$ & 62.90 & $1.58 \times 10^{10}$ &
    51.70 & $4.42 \times 10^{10}$ & 0.968 & rad.\&
    tang. vel. mass ratio 1.8 \\  
    \hline
    4a &  5 & $2.40 \times 10^{10}$ & 59.50 &  $2.40 \times 10^{10}$ &
    59.50 & $4.80 \times 10^{10}$ & 1.450 & rad.\&
    tang. vel. equal mass \\ 
    5a & 10 & $2.20 \times 10^{10}$ & 57.80 & $2.20 \times 10^{10}$ &
    57.80 & $4.40 \times 10^{10}$ & 1.372 & rad.\&
    tang. vel. equal mass \\
    6a & 20 & $2.10 \times 10^{10}$ & 56.91 & $2.10 \times 10^{10}$ &
    56.91 & $4.20 \times 10^{10}$ & 1.284 & rad.\&
    tang. vel. equal mass \\ 
    \hline
     \end{tabular}
\end{table*}

\subsection{Method}
\label{sec:method}

\citet{Jon10} investigated the minimum mass for the two DM haloes
of the satellites, by assuming they were point masses.  We use a simple
two-body integration programme modelling the system in the following
way: 
\begin{itemize}
\item Both satellites are represented by analytical, rigid
  \citet[][NFW]{Nav97} potentials.  The force on the centre of mass of 
  one halo is computed using the exact force according to its position
  with respect to the potential of the other halo.
\item To be a tightly bound pair we adopt a rigid distance criterion
  which requires that neither centre of the two haloes leaves the halo
  of the other dwarf, i.e.\ their separation is always smaller than
  the virial radius of the (smaller) halo. 
\item For each case we choose the total mass of the system {\bf and
    set up the two haloes according to their mass-ratio and
    concentration.}  Then we run the two-body code to determine if our
  distance criterion is fulfilled.  If the maximum distance is larger
  or smaller, we alter the the total mass respectively and use the
  code again.  We iterate this process until we find a maximum
  separation equal to our distance criterion. 
\end{itemize}

The reason why we choose a distance criterion instead of computing the
escape velocity (i.e. the velocity the two dwarfs need to
separate from each other to infinity) is that if we were to adopt
such a criterion, we would also include bound cases in which the
maximum separation between the two dwarfs easily exceeds their
distance to the MW.  As this would not make sense, we exclude these
solutions by imposing a very rigid distance criterion.  

Furthermore, an estimation of the tidal radii of Leo~IV and V gives
values that are similar to our distance criterion.  Using 
\begin{eqnarray}
  \label{eq:tidalrad}
  r_{\rm tidal} & \approx & \left( \frac{m_{\rm dwarf}} {3 M_{\rm
        MW}(D)} \right)^{{1/3}} \cdot D
\end{eqnarray}
with $m_{\rm dwarf} \approx 4 \times 10^{10}$~M$_{\odot}$, $M_{\rm MW}(D)
\approx 10^{12}$~M$_{\odot}$ and $D \approx 165$~kpc, we get
$40$~kpc as tidal radius.  This is slightly lower than the distance
criterion used, but is also a rather rough estimate.  However, it
shows that the distance criterion is a sensible way to restrict the
solutions.  

Hence, the choice of our distance criterion allows us to treat the
galaxy pair as isolated, i.e.\ to neglect the potential of the MW.

The simulations are always computed forward in time, starting from our
current view of the dwarfs.  We thus ascertain the next maximum
separation to assess if the two dwarfs are bound to each other now.
As we do not know where they came from, nor the details of their orbit
around the MW, we cannot predict their future fate using these models.

\subsection{Full N-body simulations}
\label{sec:sbox}

To ensure that the results are reasonable, we perform full N-body
simulations as a check of each of the 12 solutions obtained
with our simple code.  We use the particle-mesh code {\sc Superbox}
\citep{Fel00}.  It is fast and enables simulations of galaxies on
normal desktop computers.  It has two levels of higher resolution
grids, which stay focused on the simulated objects, providing high
resolution only in the areas where it is needed.  

Each object (halo) is modeled using 1,000,000 particles.  We use NFW
distributions for the haloes according to the results obtained with
the two-body code.  The haloes extend all the way to their virial
radius.    

The resolution of the grids is such that we try to keep about 15
cell-lengths per scale length, $r_{\rm sc}$, of the haloes.  A
particle-mesh code has no softening-length like a Tree-code but
previous studies \citep{spi03} showed that the length of one grid-cell
is approximately the equivalent of the softening-length.  
Furthermore the particles in a particle-mesh code are not stars or
in our case DM-particles, they rather represent tracer-particles of
the phase-space of the simulated object.  {\bf Densities are derived
  on a grid, and then a smoothed potential is calculated.}
The number of particles is chosen according to the adopted
grid-resolution to ensure smooth density distributions.  A detailed
discussion about the particle-mesh code {\sc Superbox} can be found
in \citet{Fel00}.  

To clarify we state once more, that the N-body simulations are only
used to verify the results of the simple method.  This means we
check the next maximum distance of the two galaxies.  The
simulations do not represent full-scale simulations of the past,
present and future of the two Leo galaxies.  Such simulations are
more demanding and are beyond the simple scope of this paper.

\section{Results}
\label{sec:res}

\begin{figure}
  \begin{center}
    \epsfxsize=8cm
    \epsfysize=8cm
    \epsffile{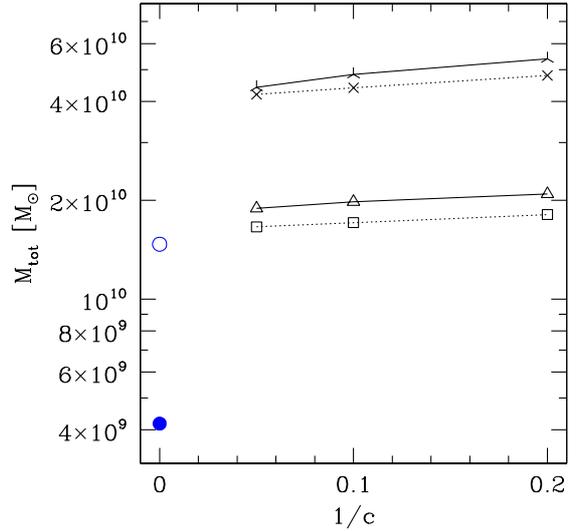}
    \caption{Minimum DM mass of the total system $M_{\rm tot}$ versus
      concentration of the haloes.  We plot $1/c$ in favour of $c$ to
      include the point-mass results at $1/c=0$ (cases 0a,b; plotted
      as open and filled circle - blue online).  Solid lines are the
      results using the mass-ratio of $1.8$. Triangles are cases with
      radial velocity only and tri-stars with additional tangential
      velocity.  Dashed lines show the results of equal mass haloes.
      Squares are radial velocity only cases and crosses have
      additional tangential velocity.} 
    \label{fig:res}
  \end{center}
\end{figure}

\subsection{Point Mass Case}
\label{sec:point}

We first recalculate the minimal bound mass assuming both galaxies are
point masses, following the methodology by \citet{Jon10}.  We assume
that the relative velocity between the two dwarfs is equal to their
escape velocity (i.e. the velocity required for the haloes to separate
to infinity).  Because point masses do not have any characteristic
radius we cannot apply any meaningful distance criterion here.

Using only the observed radial velocities (case 0a) we confirm the
total DM mass of the system obtained by \citet{Jon10}. Our result is a
factor of two lower, however, those authors used an approximation
obtained from energy arguments, while we perform the full escape
velocity calculation.  Given the wide range in possible results, as
shown later in the text, and taking the observational uncertainties
into account, a difference of a factor of two is still a very good
match. 

Furthermore, we calculate the minimum bound mass assuming the two
satellites also have tangential velocities (according to
Tab.~\ref{tab:vel}).  In this case (case 0b) the total mass required
to keep the galaxies bound is $1.47\times 10^{10}$~M$_{\odot}$. That
is, three times more DM is required to keep them bound.  While this
may seem a large mass, we refer the reader to the following sections
to put this result into context. 

The results are shown in Tab.~\ref{tab:res} in the first two lines and
in Fig.~\ref{fig:res} at $1/c = 0$.

\subsection{Two-body Integrator}
\label{sec:2halo}

\begin{table}
  \centering
  \caption{In this table we show the mass-range of the haloes of cases
    1(a)--3(a) and 4(a)--6(a) (which encompass the range of our results)
    within a radius of Leo~IV of $97$~pc, for which an observationally
    based estimate of the mass exists \citep{Sim07}}
  \label{tab:leo4}
  \begin{tabular}{ccc}
    Cases & $M^{\rm LeoIV}_{\rm DM}(r_{\rm opt}=97pc)$ &
    $M/L(r_{\rm opt})$ \\  
    & [M$_{\odot}$] & [M$_{\odot}$/L$_{\odot}$] \\
    \hline
    Simon \& Geha 2007 & $(1.4 \pm 1.5)\times 10^{6}$ & 151 \\ 
    1 -- 3 & $6.76 \times 10^{5}$ -- $4.60 \times 10^{6}$ & 37 -- 255
    \\
    1a -- 3a & $5.92 \times 10^{5}$ -- $4.03 \times 10^{6}$ & 33 --
    224 \\
    4 -- 6 & $9.31 \times 10^{5}$ -- $6.19 \times 10^{6}$ & 52 -- 344
    \\  
    4a -- 6a & $8.23 \times 10^{5}$ -- $5.58 \times 10^{6}$ & 46 --
    310 \\ \hline     
    \end{tabular}
\end{table}

We now have a look at our results, obtained by the two-body integrator
we use.  As explained in Sect.~\ref{sec:setup} the two haloes are
rigid, analytical potentials acting on the centre of mass of the other
galaxy.  Additionally, we now introduce a very strict distance
criterion of the form that neither halo centre should leave the other
halo (i.e.\ separations larger than the (smaller) virial radius).
This way we make sure that we are really dealing with a tightly bound
pair. 

The results are two-fold: Of course we see an immediate large increase
in the required minimum mass (compared to the point-mass cases) just
by introducing the rigid distance criterion.  We plot the
total mass in DM against $1/c$ in Fig.~\ref{fig:res}, to include the
point mass cases ($c=\infty$).  We see that the bound mass is larger
for lower values of the concentration.  This can be easily understood
as with higher concentrations we have more of the total mass of the
halo concentrated towards the centre and therefore the gravitational
pull on the other dwarf is larger. 

Secondly, we also find that including the additional tangential
velocity roughly triples the required mass.  While the cases with
radial velocities (cases 1--3, 1a--3a) require masses in the range of
about $1.6$--$2.1 \times 10^{10}$~M$_{\odot}$, the additional
tangential velocity increases the necessary masses up to much larger
values of $4.2$--$5.4 \times 10^{10}$~M$_{\odot}$ (cases 4--6,
4a--6a). 

As a little side-remark, we see that we need slightly less massive
haloes in the equal halo mass cases than if we adopt a mass-ratio of
$1.8$ between the two haloes.  As these differences are small compared
with the differences of the unknown concentration, and even more so
with the unknown tangential velocity, we can easily neglect them and
assume that distributing the mass differently between the two haloes
has no strong effect on our results.

The possible range of DM masses for the two galaxies spans about half
an order of magnitude.  However, given the large observational
uncertainties it is the best we can do.  The masses themselves are
rather large and taken at face value would imply that the two dwarfs
are among the most DM dominated objects in the observed universe.

\subsection{Comparison with observationally obtained data}
\label{sec:obs}

\begin{figure}
  \begin{center}
    \epsfxsize=8cm
    \epsfysize=8cm
    \epsffile{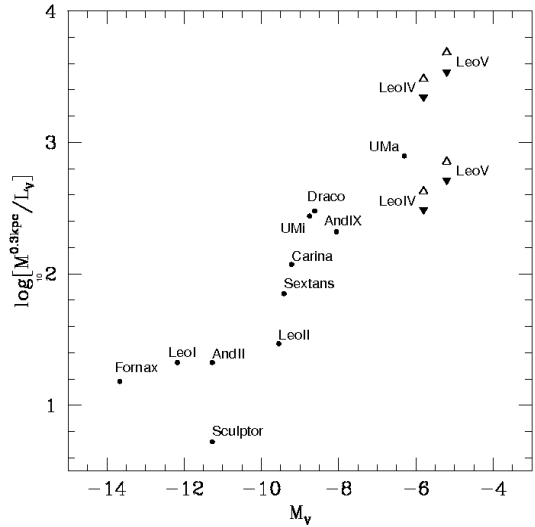}
    \caption{Mass-to-light ratios within a radius of $300$~pc.  The
      circles are the dSph galaxies of the Local Group as reported by 
      \citet{Wil06}.  For Leo~IV, the filled upside-down triangles are
      cases 1a and 3 (concentrations $c = 5$ and $c = 20$ respectively)
      with radial velocities only. The open triangles are cases 4a and
      6 ($c = 5$ and $c = 20$) with radial and tangential velocities.
      For Leo~V we plot the values of cases 1 and 3a as well as 4 and
      6a, respectively.  We use these specific cases as they span the
      whole range of our results.} 
    \label{fig:ml}
  \end{center}
\end{figure}

\begin{table}
  \centering
  \caption{Mass-to-Light ratios within a radius of $300$~pc of our
    simulations.  We adopt $V$-band magnitudes of $-5.8$ for Leo~IV
    and $-5.2$ for Leo~V.}    
  \label{tab:ml}
  \begin{tabular}{cccc}
    & & LEO IV & LEO V \\
    Case & c & Log$_{10}$($M_{\odot}$/$L_{\odot}$) &
    Log$_{10}$($M_{\odot}$/$L_{\odot}$) \\ 
    \hline
    1 &  5 & 2.544 & 2.710 \\
    2 & 10 & 2.929 & 3.092 \\
    3 & 20 & 3.343 & 3.499 \\ 
    \hline
    1a &  5 & 2.484 & 2.740 \\
    2a & 10 & 2.867 & 3.122 \\
    3a & 20 & 3.279 & 3.534 \\ 
    \hline
    4 &  5 & 2.686 & 2.853 \\
    5 & 10 & 3.067 & 3.232 \\
    6 & 20 & 3.482 & 3.641 \\ 
    \hline
    4a &  5 & 2.631 & 2.886 \\
    5a & 10 & 3.014 & 3.270 \\
    6a & 20 & 3.425 & 3.688 \\ 
    \hline
  \end{tabular}
\end{table}

\begin{figure*}
  \centering
  \epsfxsize=5.8cm
  \epsfysize=5.8cm
  \epsffile{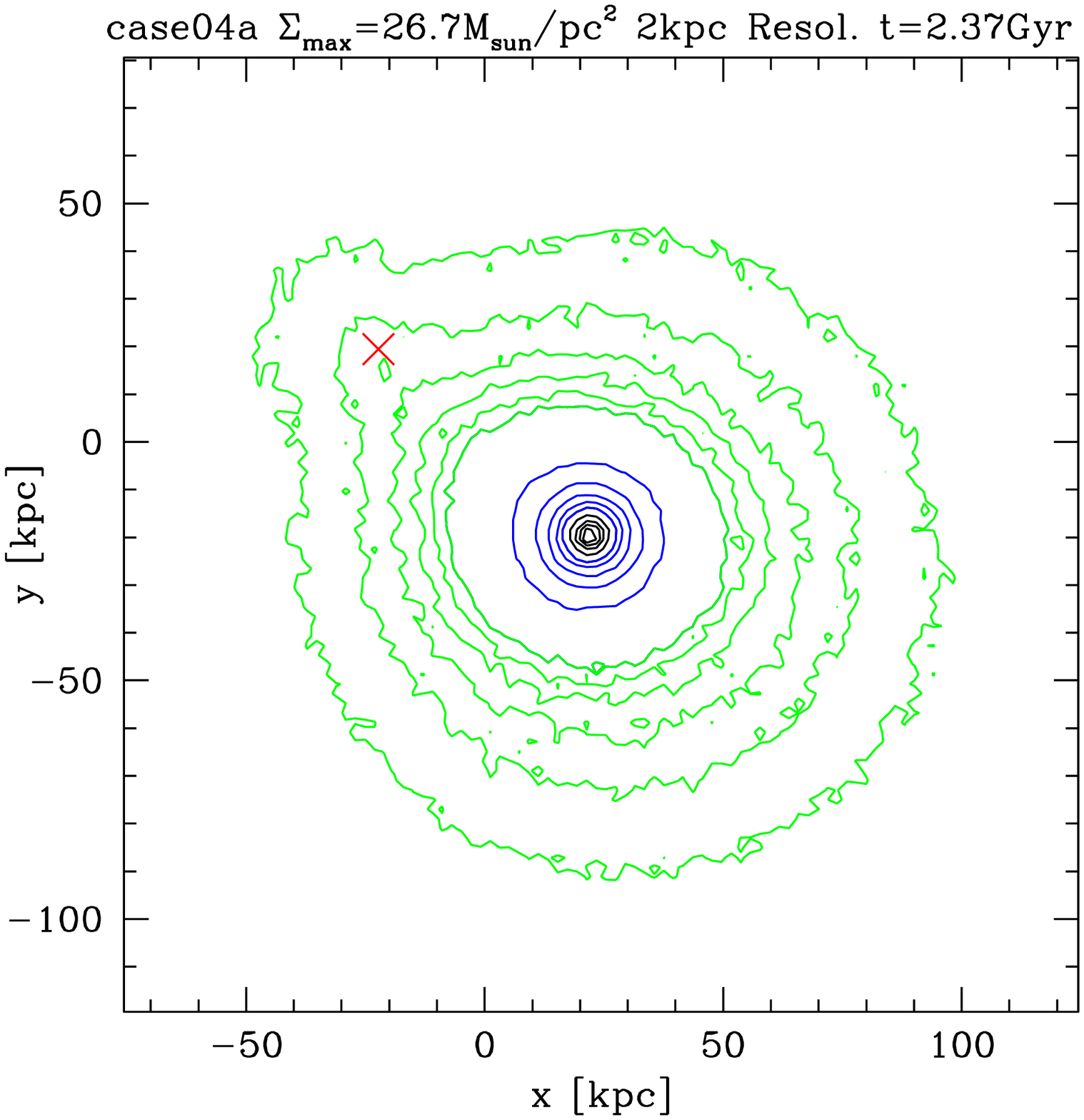}
  \epsfxsize=5.8cm
  \epsfysize=5.8cm
  \epsffile{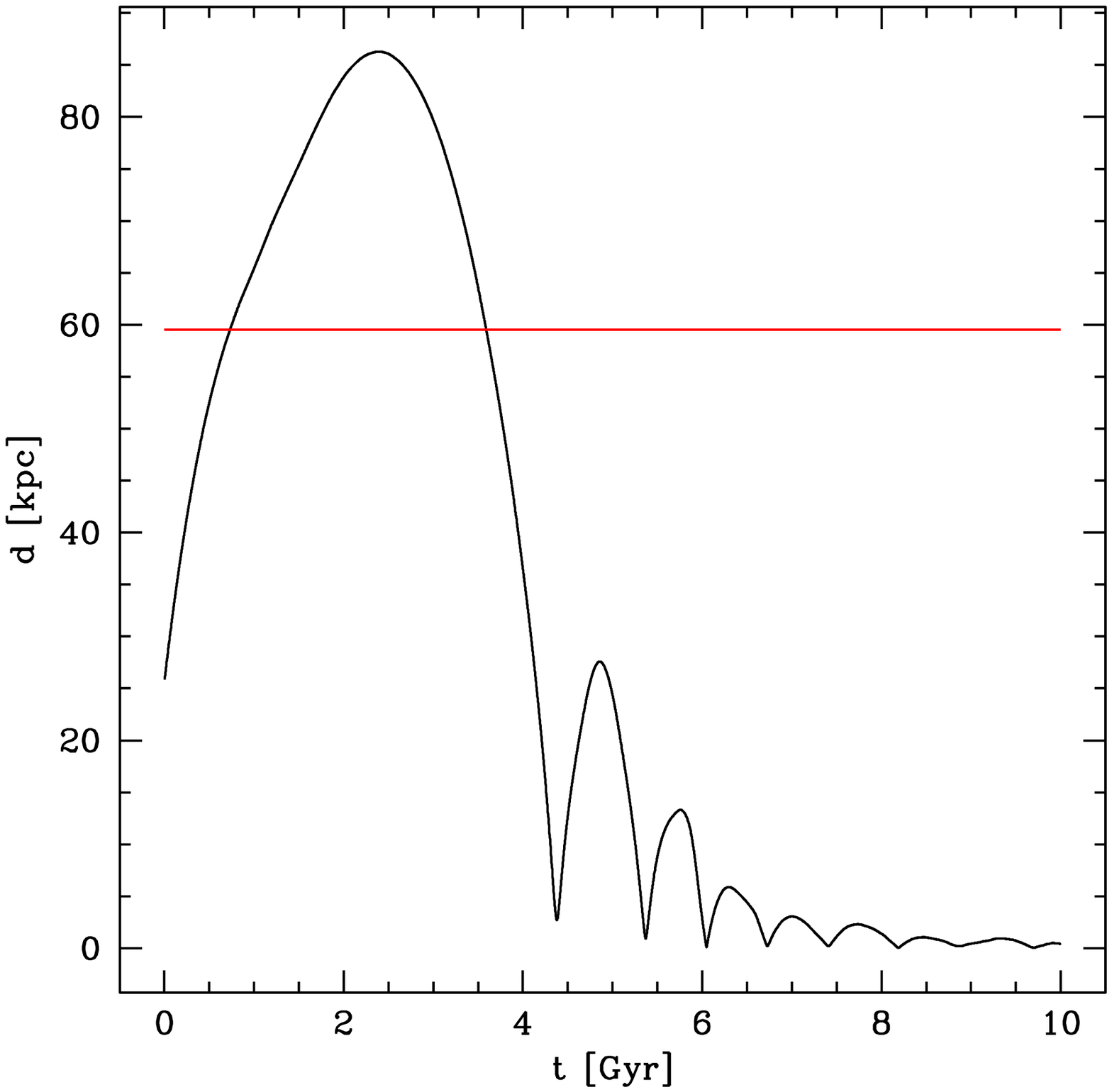}
  \epsfxsize=5.8cm
  \epsfysize=5.8cm
  \epsffile{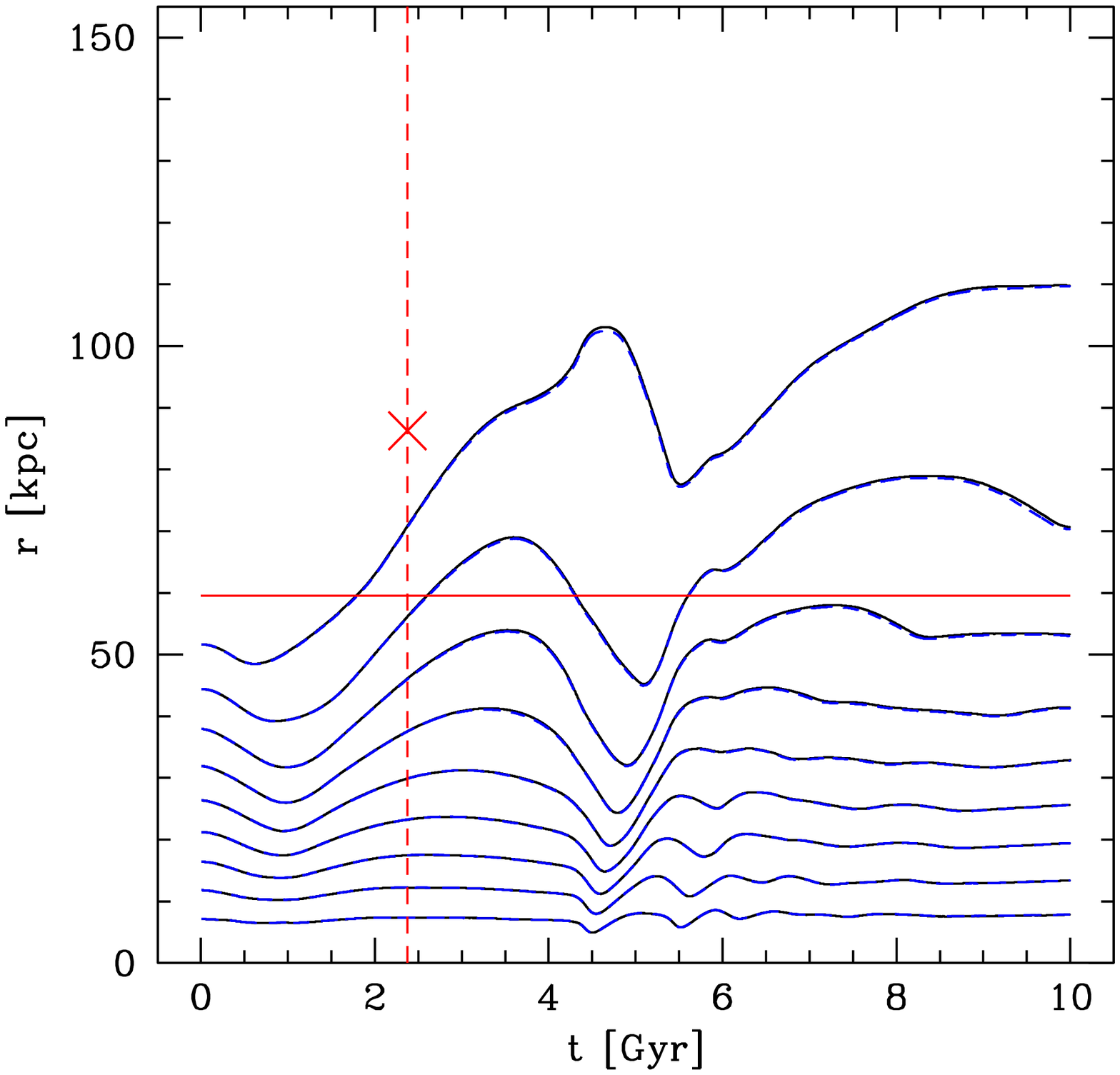}
  \caption{Results of the full N-body simulation of case 4a (the case
    with the largest discrepancy between the two-body and full N-body
    result). In the left panel the surface-density contours of the
    Leo~V halo are shown.  The (red) cross marks the position of the
    centre of Leo~IV.  The inner 5 (black) contours are between
    $26$--$10$~M$_{\odot}$\,pc$^{-2}$, and the next 5 (blue) are
    between $10$ and $1$ and the outermost 5 (green) contours are
    between $1$ and $0.1$. Middle panel: The solid (black) line shows
    the distance between the two haloes. The (red) horizontal line is
    the maximum distance from the two-body result.  The right panel
    shows the Lagrangian radii (10, 20, 30,...90~per cent of the mass)
    of the two haloes; (black) solid lines for Leo~IV and dashed
    (blue) lines for Leo~V.  The horizontal (red) solid line shows
    again the distance criterion from the simple code.  The (red) cross
    shows the maximum distance in the full simulation and vertical
    dashed (red) line marks the time of maximum separation.}
  \label{fig:tdmh}
\end{figure*}

We now put our results into context and compare them with
observations.  \citet{Sim07} measured the velocity dispersion of
Leo~IV and derived a total dynamical mass, within their optical radius
($97$~pc), of $1.4 \pm 1.5 \times 10^{6}$~M$_{\odot}$.  This implies a
M/L-ratio of $151$.  This value is quite similar to most of the other
known dSph galaxies of the MW.  We, therefore, use our results and
compute the mass of our Leo~IV haloes within the same $97$~pc.  The
resulting masses and derived M/L-ratios are shown in
Tab.~\ref{tab:leo4}.  Our M/L-ratios are in the range of $35$--$328$
and encompass the results of \citet{Sim07}.  Furthermore, should the
results of \citet{Sim07}, which are based on very few stars, prove to
be correct, our results mean that we can rule out DM haloes with high
concentrations (i.e. $c=20$).

\citet{Wal09a} report a central velocity dispersion of $\sigma =
2.4^{+2.4}_{-1.4}$~km\,s$^{-1}$ based on five stars for Leo~V.  Using
this value they calculate a dynamical mass, within an adopted $r_{\rm
  h} = 67.4$~pc, of $3.3^{+9.1}_{-2.5} \times 10^{5}$~M$_{\odot}$.
Calculating the mass within this radius in our models gives a range
from $2.7 \times 10^{5}$~M$_{\odot}$ to $2.5 \times
10^{6}$~M$_{\odot}$, again encompassing the results derived from
observations by \citet{Wal09a}. 

Another way to compare our results with observations is by computing
the total mass within a `standard' radius of $300$~pc, as adopted by
\citet{Wal09b}.  We can then infer the M/L-ratio inside this radius
and compare our results with the observationally derived results
reported by \citet{Wil06}.  Given the low luminous masses of the two
dwarfs, our results point to M/L-ratios in the order of
$\log_{10}(M/L) = 2.5$--$3.7$ (see also Tab.~\ref{tab:ml}).  Despite
being quite high, plotting these values together with observationally
derived values of other dwarfs (Fig.~\ref{fig:ml}) we see that they
follow the general trend of higher M/L-ratios with lower luminous
masses.  In fact, if we fit a line through the observational results,
our values would intersect that linear fit.

This means, not only do our results follow and confirm the observed
trend of the known MW dSphs, we can further conclude that the two
satellites do not need unreasonable amounts of DM to form a bound
pair. 

\subsection{Comparison to N-body simulations}
\label{sec:comp}

Since we now have the results of all our cases, we have to make sure
that they still hold if we resimulate them with a full N-body code.
Of course the full N-body simulations will differ significantly from
the ones above, we simply want to know if our conclusions remain
valid. 

In the full N-body simulations, the two live haloes are interacting
with each other.  They experience dynamical friction which shrinks
their orbits around each other until they finally merge.  This cannot
be reproduced by the simple code but we can determine if the next
maximum separation of the orbit is smaller than the extent of the
other halo, as our distance criterion in the simple simulations
requires. 

What we find is that the results differ by a few per cent
(max. 6.5\%) in the simulations using a mass ratio of $1.8$ between
the haloes.  The haloes get slowed down and, therefore, turn
around at a smaller separation.

If the haloes have the same mass and we only adopt radial velocities,
the orbit overshoots the maximum distance by approximately 20~per
cent.  If we add the additional tangential velocities, our restricted
results differ by about 28--45~per cent, in the sense that the
maximum separation is larger in   the `full' simulation than in the
restricted one.  This may seem odd given that dynamical friction
should act in the opposite direction, however, there are other
mechanisms at work.  We see an expansion of the haloes as orbital 
energy is transformed into internal energy, furthermore, we see that
the two haloes get deformed -- particles from one halo get dragged
along by the gravitational force of the other. We give the ratio of
the maximum separations between the full and restricted simulations in
the second to last column of Tab.~\ref{tab:res} (labeled `ratio').

We plot, in Fig.~\ref{fig:tdmh}, the results of case 4a, the case with
the largest discrepancy in separation between the two haloes compared
with the restricted prediction.  In the left panel we see the contours
of the Leo~V halo at the time of maximum separation, with the cross
marking the position of the centre of Leo~IV.  We see that the
contours are slightly elongated towards the other halo and that they
show a clear deformation.  This deformation is caused by the
gravitational pull of the other halo, which has dragged particles of
the dwarf towards it. 

In the middle panel we see the large discrepancy between the distance 
criterion (horizontal line) and the actual first maximum separation of
the orbit.  But as the total mass of a NFW profile only increases with
the logarithm of the radius, even a large discrepancy in radius as in
our case 4a amounts only to a few per cent error in the mass of the
halo.  Given the fact that our results span almost an order of
magnitude, and the large uncertainties from the observations (luminous
mass, distance, etc.), we claim that the results of the restricted
code are verified. 

Finally, the right panel of Fig.~\ref{fig:tdmh} shows the Lagrangian
radii of Leo~IV (solid lines) and Leo~V (dashed lines).  We see that
the interaction of the two haloes causes the Lagrangian radii to
expand. At the time of the maximum separation (marked with a cross)
the halo of Leo~IV is just outside the $90$~\% mass-radius and,
as shown in the left panel, is still within the expanded and
deformed halo of Leo~V.  In some sense, this matches the original
distance criterion, which said that neither halo-centre should 
leave the other halo.

\section{Discussion \& Conclusions}
\label{sec:conc}

We have presented possible scenarios for a twin system consisting of
the faint dwarf spheroidal galaxies Leo~IV and Leo~V.  The simulations
were performed using a simple two-body code to rapidly find the
solutions in the vast parameter space and the results were verified
using a full N-body code.  From this we find the minimum DM masses
required for the two galaxies to form a tightly bound pair.

The parameter space is restricted by assuming two independent DM
haloes orbiting each other.  Two perfectly shaped haloes would only be
seen before the first close passage.  {\bf This is a strong
  simplification of the real geometry of the problem.  But as our
  results show (i.e.\ the comparison with the real N-body simulations)
  the resulting error of this simplification is in the order of 5-20\%
  and therefore much smaller than the mass-range of our results,
  stemming from e.g.\ the unknown tangential velocity.}

A further restriction is the maximum distance criterion we adopt.  We
find this criterion sensible given the satellites' large distances
from the MW.  Smaller maximum separations would lead to higher
required masses for the system to be bound.  Larger separations
would lead to lower masses, but since the expected tidal radius of
the system (with respect to the gravitational force of the MW) is of
the order of our distance criterion, we feel confident with our
choice. 
 
As our distance criterion is of the order of the tidal radius, we
are able to simplify even further and treat the system of the two
dwarfs as isolated (i.e.\ we do not simulate the potential of the
MW).  As our aim is to determine whether the two galaxies are bound
now (and make no predictions about their future or past), we do not
need to take their orbit around the MW into account.

Regarding the relative velocity we adopt two cases.  In one the
restriction is that the measured difference in radial velocity is the
only relative velocity the dwarfs have.  In the other case the two
satellites are given an additional tangential velocity of the same
magnitude as the radial velocity. 

We also adopt two mass ratios.  First, mass-follows-light, i.e.\ the
two haloes have a mass ratio of $1.8$ like the luminous components.
Second, mimicking {\bf the fact (claimed by \citet{Wal09b}) that
  almost all dSph galaxies reside in DM haloes of the same minimum
  mass (i.e\ a minimum halo mass to carry a luminous component)}, a
mass ratio of $1.0$.  Moreover, to span the full range of proposed
concentrations for dwarf galaxy dark haloes we take three values for
the concentration into account $c = 5, 10, 20$. 

If we assume that the bound system consists of two DM haloes orbiting
each other, we infer masses of about $1.7 - 2.1 \times
10^{10}$~M$_{\odot}$ for the whole system.  If we add an additional
tangential velocity, which cannot be observationally verified, we
obtain $\approx 4.2 - 5.4 \times 10^{10}$~M$_{\odot}$.  These are
indeed very high masses for the two faint satellites and would put
them amongst the most DM dominated objects known.  Still, these
results do not infer that the scenario is impossible. 

Another point to take away is that if we add the same amount of
relative velocity tangentially, i.e.\ increasing the total relative
velocity by a factor $\sqrt(2)$, the required mass more than doubles.
This shows quite a strong dependence on the relative velocity.  Still,
if we double the tangential velocity the mass would vary within an
order of magnitude, an uncertainty we find in our results anyway. 

We compute the DM mass within the adopted optical radius of
\citet{Sim07} for Leo~IV (i.e.\ 97~pc) and find that the M/L-ratios we
obtain span the observationally (measured velocity dispersion) derived
results.  Our results are also in agreement with the measured velocity
dispersion of Leo~V \citep{Wal09a} and the inferred dynamical mass. 
Taking the observations at face value, our results could,
therefore, restrict the possible concentrations of the real DM haloes,
once we know their relative tangential velocity.

Furthermore, we checked our results against the trend for dSph
galaxies published by \citet{Wil06}.  They give the M/L-ratios within
a radius of $300$~pc \citep[also seen in][]{Wal09b}.  Our simulations
predict M/L-ratios, using the same radius, in the range
$\log_{10}{M/L} = 2.5$--$3.7$.  These values are high but encompass
the predictions for faint dSph galaxies, if we extrapolate the known
values to the magnitudes of the Leos.

Comparing the results of our two-body code with full N-body
simulations we find differences in the maximum separations of only a
few per cent in most of the simulations.  Only the simulations with
equal mass haloes and additional tangential velocity have rather large
discrepancies.  Because the mass of an NFW halo increases with radius,
proportional to $\ln(r)$, the uncertainty in the masses is much lower.
The simple integration programme used cannot predict any deformations
of the haloes due to their mutual interactions.  In the full
simulations we see those deformations and, even though the initial
distance criterion is not fulfilled anymore, the haloes still stay
within the deformations of the other.  In that sense the distance
criterion is still obeyed.

Our final remark is that we wanted to search for the necessary total
dark matter mass of the pair of satellites to ensure that they are
bound to each other.  Even though the comparison between the simple
code and the full N-body results deviate somewhat from our distance
criterion, they do not change the conclusions of the simulations.  A
bound pair in the restricted case is still a bound pair in the full
simulations.  Just by looking at our different cases (i.e.\ radial
velocity only or radial plus tangential velocity) our results differ
by about half an order of magnitude in total mass.  In that respect a
mass uncertainty of even 20--30~per cent does not change the
conclusions of this paper, nor would it alter the inferred M/L-ratios
significantly. 

Summing up, assuming that the two Leos do, in fact, consist of a
tightly bound pair, we find their inferred dark matter masses to be
high but still within reasonable values.  Therefore, it is possible
that the two galaxies form a bound pair, making them an ultra-faint
counterpart of the Magellanic Clouds. 

\begin{acknowledgements}
  MF acknowledges funding through FONDECYT grant 1095092 and BASAL. RS 
  is funded through a Comite Mixto grant. 
\end{acknowledgements}

\label{lastpage}


\begin{thebibliography}{XXX99}

\bibitem[\protect\citeauthoryear{Belokurov et al.}{2006}]{Bel06}
  Belokurov V., et al. 2006, ApJ, 647, L111

\bibitem[\protect\citeauthoryear{Belokurov et al.}{2007}]{Bel07}
  Belokurov V., et al. 2007, ApJ, 654, 897

\bibitem[\protect\citeauthoryear{Belokurov et al.}{2008}]{Bel08}
  Belokurov V., et al. 2008, ApJ, 686, L83

\bibitem[\protect\citeauthoryear{Boylan-Kolchin et al.}{2009}]{Boy09}
  Boylan-Kolchin M., Springel V., White S.D., Jenkins A., Lemson
  G. 2009, MNRAS, 398, 1150

\bibitem[\protect\citeauthoryear{Coleman et al.}{2007}]{Col07}
  Coleman M.G., et al. 2007, ApJ, 668, L43

\bibitem[\protect\citeauthoryear{Fellhauer et al.}{2000}]{Fel00}
  Fellhauer M., Kroupa P., Baumgardt H., Bien R., Boily C., Spurzem
  R., Wassmer N. 2000, NewAst., Vol.~5, No.~6, 305

\bibitem[\protect\citeauthoryear{Fellhauer et al.}{2006}]{Fel06}
  Fellhauer M., et al. 2006, MNRAS, 385, 1095

\bibitem[\protect\citeauthoryear{Fellhauer et al.}{2007}]{Fel07}
  Fellhauer M., et al. 2007, MNRAS, 375, 1171

\bibitem[\protect\citeauthoryear{Fellhauer et al.}{2008}]{Fel08}
  Fellhauer M., et al. 2008, MNRAS, 385, 1095

\bibitem[\protect\citeauthoryear{Geha et al.}{2009}]{Geh09}
  Geha M., Willman B., Simon J.D., Strigari L.E., Kirby E.N., Law
  D.R., Strader J. 2009, ApJ, 692, 1464

\bibitem[\protect\citeauthoryear{Jin et al.}{2012}]{Jin12}
  Jin S., Martin N., de Jong J., Conn B., H.-W., Irwin M. 2012 to
  appear in the proceedings of the Subaru conference on Galactic
  Archaeology, Shuzenji, Japan (Nov. 1-4 2011), arXiv1201.5399 

\bibitem[\protect\citeauthoryear{de Jong et al.}{2008}]{Jon08}
  de Jong J.T.A., et al. 2008, ApJ, 680, 1112

\bibitem[\protect\citeauthoryear{de Jong et al.}{2010}]{Jon10}
  de Jong J.T.A., Martin N.F., Rix H.-W., Smith K.W., Jin S.,
  Macci\`{o} A.V. 2010, ApJ, 710, 1664

\bibitem[\protect\citeauthoryear{Kirby et al.}{2008}]{Kir08}
  Kirby E.N., Simon J.D., Geha M., Guhathakurta P., Frebel A. 2008,
  ApJ, 685, L43

\bibitem[\protect\citeauthoryear{Klypin et al.}{1999}]{Kly99}
  Klypin A., Kravtsov A.V., Valanzuela O., Prada F. 1999, ApJ, 522, 82 

\bibitem[\protect\citeauthoryear{Koch et al.}{2009}]{Koc09}
  Koch A., et al. 2009, ApJ, 690, 453

\bibitem[\protect\citeauthoryear{Koposov et al.}{2008}]{Kop08}
  Koposov S., et al. 2008, ApJ, 686, 279

\bibitem[\protect\citeauthoryear{Kuhlen et al.}{2008}]{Kuh08}
  Kuhlen M., Diemand J., Madau P., Zemp M. 2008, Journal of Physics:
  Conference Series, Vol.\ 125, Issue 1, p.~1

\bibitem[\protect\citeauthoryear{Lokas \& Mamon}{2001}]{Lok01}	
  Lokas E.L., Mamon G.A. 2001, MNRAS, 321, 155

\bibitem[\protect\citeauthoryear{Macci\`{o} et al.}{2009}]{Mac09}
  Macci\`{o} A.V., Dutton A.A., van den Bosch F.C. 2009, MNRAS, 391,
  1940 

\bibitem[\protect\citeauthoryear{Macci\`{o} et al.}{2010}]{Mac10}
  Macci\`{o} A.V., Kang X., Fontanot F., Somerville R.S., Koposov S.,
  Monaco P. 2010, MNRAS, 402, 1995

\bibitem[\protect\citeauthoryear{Metz, Kroupa \& Libeskind}{Metz et
    al.}{2008}]{Met08} 
  Metz M., Kroupa P., Libeskind N.I. 2008, ApJ, 680, 287

\bibitem[\protect\citeauthoryear{Metz, Kroupa \& Jerjen}{Metz et
    al.}{2009}]{Met09}
  Metz M., Kroupa P., Jerjen H. 2009, MNRAS, 394, 1529

\bibitem[\protect\citeauthoryear{Moore et al.}{1999}]{Moo99}
  Moore B., Ghigna S., Governato F., Lake G., Quinn T., Stadel J.,
  Tozzi P. 1999, ApJ, 524, L19

\bibitem[\protect\citeauthoryear{Moretti et al.}{2009}]{Mor09}
  Moretti M.I., et al. 2009, ApJ, 699, L125

\bibitem[\protect\citeauthoryear{Mu\~{n}oz et al.}{2008}]{Mun08}
  Mu\~{n}oz R.R., Majewski S.J., Johnston K.V. 2008, ApJ, 679, 346

\bibitem[\protect\citeauthoryear{Navarro, Frenk \&
    White}{1997}]{Nav97}
  Navarro J.F., Frenk C.S., White S.D.M. 1997, ApJ, 490, 493

\bibitem[\protect\citeauthoryear{Pawlowski et al.}{2011}]{Paw11}
  Pawlowski M.S., Kroupa P., de Boer K.S. 2011, A\&A, 523, 118

\bibitem[\protect\citeauthoryear{Sand et al.}{2009}]{San09}
  Sand D.J., Olszewski E.W., Willman B., Zaritsky D., Seth A., Harris
  J., Piatek S., Saha A. 2009, ApJ, 704, 898

\bibitem[\protect\citeauthoryear{Sand et al.}{2010}]{San10}
  Sand D.J., Seth A., Olszewski E.W., Willman B., Zaritzky D.,
  Kallivayalil N. 2010, ApJ, 718, 530

\bibitem[\protect\citeauthoryear{Simon \& Geha}{2007}]{Sim07}
  Simon J.D., Geha M. 2007, ApJ, 670, 313

\bibitem[\protect\citeauthoryear{Simon et al.}{2010}]{Sim10}
  Simon J.D., Frebel A., McWiliam A., Kirby E.N., Thompson I.B. 2010,
  ApJ, 716, 446

\bibitem[\protect\citeauthoryear{Spinnato, Fellhauer \& Portegies
    Zwart}{Spinnato et al.}{2003}]{spi03}
  Spinnato P.F., Fellhauer M., Portegies Zwart S.F. (2003), MNRAS,
  344, 22

\bibitem[\protect\citeauthoryear{Walker et al.}{2009a}]{Wal09a}
  Walker M.G., Belokurov V., Evans N.W., Irwin M.J., Mateo M.,
  Olszewski E.W., Gilmore G. 2009, ApJ, 694, L144
	
\bibitem[\protect\citeauthoryear{Walker et al.}{2009b}]{Wal09b}	
  Walker M.G., Mateo M., Olszewski E.W., Penarrubia J., Evans, N.W.,
  Gilmore G. 2009, ApJ, 704, 1274

\bibitem[\protect\citeauthoryear{Walsh, Jerjen \& Willman}{Walsh et
    al.}{2007}]{Wal07} 
  Walsh S.M., Jerjen H., Willman B. 2007, ApJ, 662, L83

\bibitem[\protect\citeauthoryear{Willman et al.}{2005}]{Wil05}
  Willman B., et al. 2005, ApJ, 626, L85

\bibitem[\protect\citeauthoryear{Wilkinson et al.}{2006}]{Wil06}
  Wilkinson M.I., Kleyna J.T., Gilmore G.F., Evans N.W., Koch A.,
  Grebel E.K., Wyse R.F.G., Harbeck D.R. 2006, Msngr, 124, 25

\bibitem[\protect\citeauthoryear{Zucker et al.}{2006a}]{Zuc06a}
  Zucker D.B., et al. 2006, ApJ, 643, L103

\bibitem[\protect\citeauthoryear{Zucker et al.}{2006b}]{Zuc06b}
  Zucker D.B., et al. 2006, ApJ, 650, L41

\end{thebibliography}
\end{document}